\documentclass[sigconf]{acmart}

\usepackage{url}
\usepackage{multirow}
\usepackage{multicol}
\usepackage{enumitem}
\usepackage{color,soul} 
\usepackage{listings}
\usepackage{adjustbox}

\newcommand{\filipe}[1]{\textcolor{blue}{{\it [Filipe says: #1]}}}

\newcommand{\boyuan}[1]{\textcolor{purple}{{\it [Nemo says: #1]}}}

\AtBeginDocument{%
  \providecommand\BibTeX{{%
    \normalfont B\kern-0.5em{\scshape i\kern-0.25em b}\kern-0.8em\TeX}}}

\setcopyright{acmcopyright}
\copyrightyear{2022}
\acmYear{2022}
\acmDOI{10.1145/3510457.3513050}

\acmConference[ICSE-SEIP '22]{44nd International
Conference on Software Engineering: Software Engineering in Practice}{May 21--29,
2022}{Pittsburgh, PA, USA}
\acmBooktitle{44nd International Conference on Software Engineering: Software
Engineering in Practice (ICSE-SEIP '22), May 21--29, 2022, Pittsburgh, PA, USA}
\acmPrice{15.00}
\acmISBN{978-1-4503-9226-6/22/05}

\begin{document}

\title{Towards Build Verifiability for Java-based Systems}

\author{Jiawen Xiong$^1$, Yong Shi$^2$, Boyuan Chen$^3$, Filipe R. Cogo$^4$, Zhen Ming (Jack) Jiang$^5$}
\affiliation{%
	\institution{Huawei China$^{1,2}$, Huawei Canada$^{3,4}$, York University$^5$}
	\country{Shenzhen, China$^{1,2}$, Kinston, Canada$^{3,4}$, Toronto, Canada$^5$}}
\email{{xiongjiawen, young.shi, boyuan.chen1, filipe.roseiro.cogo1}@huawei.com, zmjiang@eecs.yorku.ca}
\renewcommand{\shortauthors}{Xiong and Shi, et al.}

\begin{abstract}
  Build verifiability refers to the property that the build of a software system can be verified by independent third parties and it is crucial for the trustworthiness of a software system. 
  Various efforts towards build verifiability have been made to \textsf{C/C++}-based systems, yet the techniques for \textsf{Java}-based systems are not systematic and are often specific to a particular build tool (e.g., \textsf{Maven}). In this study, we present a systematic approach towards build verifiability on \textsf{Java}-based systems. 
  Our approach consists of three parts: a unified build process, a tool that dynamically controls non-determinism during the build process, and another tool that eliminates non-equivalences by post-processing the build artifacts.
  We apply our approach on 46 unverified open source projects from \textsf{Reproducible Central} and 13 open source projects that are widely used by \textsf{Huawei} commercial products. As a result, 91\% of the unverified \textsf{Reproducible Central} projects and 100\% of the commercially adopted OSS projects are successfully verified with our approach. In addition, based on our experience in analyzing thousands of builds for both commercial and open source \textsf{Java}-based systems, we present 14 patterns that introduce non-equivalences in generated build artifacts and their respective mitigation strategies. Among these patterns, 11 (78\%) are unique for \textsf{Java}-based system, whereas the remaining 3 (22\%) are common with \textsf{C/C++}-based systems. The approach and the findings of this paper are useful for both practitioners and researchers who are interested in build verifiability.

\end{abstract}

%



\maketitle

\section{Introduction}

\textsf{Java} is one of the most prominent programming languages in the software industry, ranked third in the TIOBE index~\cite{tiobe:2021}. Given the popularity of \textsf{Java}, both industry and open source initiatives are actively researching forms of improving the security of applications written in this programming language. Build verifiability is an important security property that ensures correspondence between the source code and the deliverable packages that are distributed to final users. Build verifiability is paramount for both commercial and Open Source Software (OSS) systems and can potentially prevent incidents such as software supply chain attacks~\cite{sonatype_suppchain_attack,preventing_suppchain_attacks_solarwind, marc:backstabber_knife:2020,vu_duc:using_source_repo_supp_attack:2020}, which silently injects malicious code into a distributed package during the build process~\cite{thompson:trusting_trust:1984}. Before we can consider a build setup as a \emph{verifiable build}~\cite{Carnavalet:2014}, one of the following two properties needs to be satisfied by the generated deliverable packages: (1) generated packages by two build instances are always equivalent (i.e., have the same contents), or (2) the technical details behind occasional non-equivalences in the built packages can be explained (e.g., using Name Space Layout Randomization~\cite{nslr:2017} to defend against code injection attacks). When all deliverable packages satisfy the first property, we consider the build setup as a \emph{reproducible build}~\cite{ReproBuild:2021}.

Producing a verifiable build is not trivial due to sources of non-determinism present in the build toolchain, the build environment, or the design of a software system. Prior research proposed different approaches towards producing verifiable builds of \textsf{C}/\textsf{C++}-based systems~\cite{leija:2020:repro_containers,Ren:2018,Ren:2019}. In particular, our prior work~\cite{ShiTSE2021} proposed a unified process and a toolkit to produce verifiable builds for \textsf{C/C++}-based large-scale industrial systems. Our unified process encompasses a catalog of remediation strategies that is periodically updated whenever new sources of non-determinism are identified and mitigated. We leverage the following three different mitigation strategies: (1) \emph{controlling}, which intercept non-deterministic build instructions at runtime and returns pre-defined values~\cite{leija:2020:repro_containers}; (2) \emph{remediation}, which modifies source code and build scripts to mitigate sources of non-determinism~\cite{Ren:2018}; and (3) \emph{interpretation}, which provides a traceable explanation of eventual non-equivalences in the build artifacts that are introduced by design~\cite{Carnavalet:2014}. Our approach has been checked by an independent auditing organization for compliance~\cite{HCSEC:2021} and is currently used by hundreds of systems within \textsf{Huawei}. However, the aforementioned existing approaches cannot be directly applied to \textsf{Java}-based systems due to the following challenges:

\begin{itemize}[leftmargin=0.5cm]

\item \textbf{Distinct sources of non-determinism.} The sources of non-determinism that cause non-equivalences in \textsf{Java} packages can be different from those of \textsf{C/C++} packages. There are no prior investigations on the sources of non-determinism that are exclusively related to \textsf{Java} systems. For example, some specific sources of non-determinism stem from how the compilation mechanism of \textsf{JavaDoc}~\cite{java_doc_compilation:2021} and \textsf{JSP}~\cite{jsp_compilation:2021} files works. Similarly, it is unknown if there are any common sources of non-determinism between \textsf{Java}- and \textsf{C/C++}-based systems. 

\item \textbf{Distinct mitigation approaches.} The approaches to mitigating the sources of non-determinism are different between \textsf{Java} and \textsf{C/C++}-based systems. For example, the controlling mechanism used in \textsf{C/C++}-based systems dynamically intercepts non-deterministic build instructions at the kernel level (e.g., using \texttt{LD\_PRELOAD} hooks~\cite{ShiTSE2021}). These mechanisms cannot work directly on \textsf{Java}-based systems, as the non-deterministic build instructions need to be intercepted at the \textsf{JVM} level. As a result, new mitigation approaches are needed for \textsf{Java}-based systems.

\item \textbf{Distinct build mechanisms.} \textsf{Java}-based systems have different build mechanisms compared to \textsf{C/C++}-based systems. Many \textsf{Java}-based systems are built by automated build tools (e.g., \textsf{Maven} and \textsf{Gradle}), which either use pre-compiled libraries locally or automatically retrieve them from central remote repositories (e.g., \textsf{Maven Central}). In addition, a \textsf{Java} package should, in principle, run in any platform with an installed \textsf{JVM} instance. This process is different from the build of \textsf{C/C++} systems whose libraries are platform-dependent and typically stored in local repositories. Therefore, we also need to consider additional aspects when analyzing and improving build verifiability for  \textsf{Java}-based systems.
\end{itemize}

Build verifiability of \textsf{Java}-based systems is supported by associated plugins with each build tool. However, these plugins require individual installation and configuration for each system and are only able to mitigate a limited set of sources of non-determinism. To tackle these challenges, in this paper, we propose a new approach to systematically diagnosing and automatically mitigating the sources of non-determinism during the build of a \textsf{Java}-based system. The automatic mitigation leverages dynamic bytecode instrumentation~\cite{java_inst_api:2021} to control the non-deterministic build instructions (a.k.a., the controlling mechanism). We further improve the interpretation mechanisms such that we not only explain the non-equivalences but also demonstrates their effect through a post-processing step. We have applied our approach on 46 projects from Reproducible Central~\cite{mavenrebuild} and 13 OSS projects that are often used by commercial applications from \textsf{Huawei}. The build from 55 (93.2\%) of the projects can now be fully verified, compared to 0 previously. The contributions of our paper are the following:

\begin{enumerate}[leftmargin=0.5cm]
    \item This is the first study that systematically investigates build verifiability for Java-based systems. Through our experience in verifying the deliverable packages of thousands of Java-based projects, we have derived a set of root causes and their associated mitigation strategies.
    \item Compared with existing approaches to producing verifiable builds for \textsf{Java}-based applications, our approach is shown: a) to mitigate sources of non-determinism that are not mitigated by existing approaches, b) to prevent the modification of existing build setups or the integration and configuration of plugins, c) to affect only specific fields and methods of specific classes, d) to integrate seamlessly with the most popularly adopted \textsf{Java} build tools, and e) to extend effortlessly to mitigate new sources of non-determinism.
    \item We report 14 patterns that yield non-deterministic build instructions in \textsf{Java}-based systems and their associated mitigation strategies. While comparing against previously reported patterns in \textsf{C/C++}-based systems, 11 patterns are new and unique for \textsf{Java}-based systems. 
\end{enumerate}

\noindent \textbf{Paper organization:} Section~\ref{sec:background-motivation} presents the motivation and background material of our paper. Section~\ref{sec:approach} presents our approach to produce verifiable builds of \textsf{Java}-based systems. Section~\ref{sec:evaluation} presents the case study results. Section~\ref{sec:discussion} discusses the results of our case study. 
Section~\ref{sec:threats-to-validity} presents the threats to the validity of our paper. Finally, Section~\ref{sec:conclusion} presents our conclusions.
\section{Background and Related Works}
\label{sec:background-motivation}

In this section, we present how \textsf{Java}-based systems are typically built (Section~\ref{sec:background-motivation:subsec:build-java-systems}), the existing approaches to producing verifiable builds (Section~\ref{sec:background-motivation:subsec:existing-approaches-verifiable-build}), and how verifiable builds are currently produced for \textsf{Java}-based systems (Section~\ref{sec:background:subsec:verifiable-builds-java-systems}).

\subsection{The build of Java-based systems}
\label{sec:background-motivation:subsec:build-java-systems}

\begin{figure}
    \includegraphics[scale=0.4]{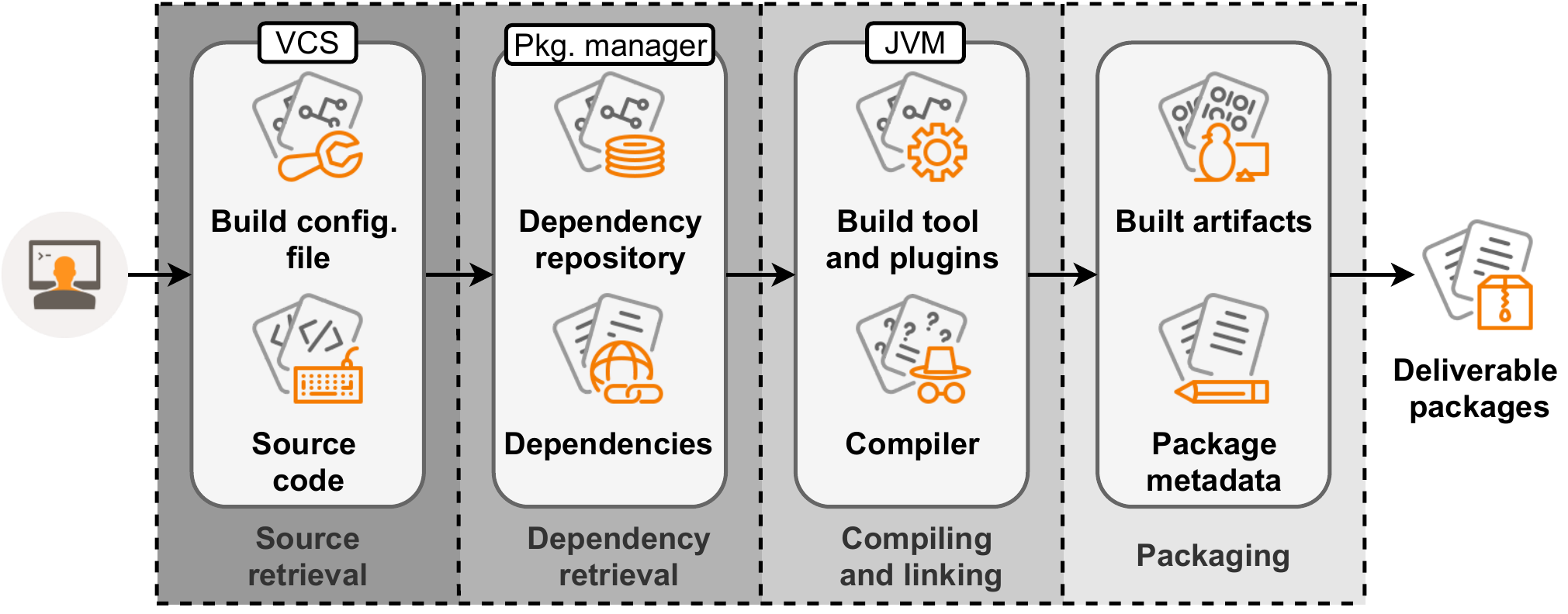}
    \centering
    \caption{The build process of \textsf{Java}-based systems.}
    \label{fig:java-build-process}
\end{figure}

The build of a \textsf{Java}-based system encompasses the four general phases as shown in Figure~\ref{fig:java-build-process}. We explain each of the general build phases below by comparing against the build process for \textsf{C/C++}-based systems.

\noindent \textbf{Source retrieval.} During the \emph{source retrieval} phase, the \emph{build configuration file} and the \emph{source code} are fetched from a Version Control System (VCS). This process is typically supported by automated build tools and is similar in both \textsf{Java}- and \textsf{C/C++}-based systems, although the automated build tools are different depending on the language. The build configuration file is responsible for setting up access to \emph{dependency repositories}, determining which \emph{dependencies} are retrieved and linked during a build process, customizing the behaviour of the \emph{build tool} and its associated \emph{plugins}, and configuring options for the \emph{compiler}.

\noindent \textbf{Dependency retrieval.} In \textsf{Java}, it is common to retrieve dependencies (e.g., \texttt{jar} files) from either local or remote (e.g., \textsf{Maven Central}) \emph{dependency repositories}. Dependencies are typically retrieved by the associated package manager with the build tool. The functionalities provided by library packages are directly reused by the built application. In contrast, in \textsf{C/C++}-based systems, dependencies are typically retrieved from local repositories of shared libraries.

\noindent \textbf{Compiling and linking.} The next phase is automatically supported by build tools and plugins and is broken down into two steps: (1) The \textsf{compiling} step compiles the source files that are retrieved in the \emph{source retrieval} phase. (2) The \textsf{linking} step binds the compiled artifacts with the obtained dependencies in the \emph{dependency retrieval} phase to produce a set of executable files or instructions living inside runtime environments (e.g., \textsf{JVM}). The output of these two steps is one or more \emph{built artifacts} (called \texttt{class files} in \textsf{Java} systems and \texttt{object files} in \textsf{C/C++} systems) that are used as input to the next phase.
In \textsf{Java}, the linking process is performed by the \textsf{JVM}. Most of the major \textsf{Java} build tools (e.g., \textsf{Ant}, \textsf{Maven}, and \textsf{Gradle}) run over the \textsf{JVM}, as they are \textsf{Java} applications themselves. Therefore, any sources of non-determinism that stem from the \textsf{JVM} also affect the \textsf{Java} build tools. 

\noindent \textbf{Packaging.} In the last phase, the built artifacts from the previous phase (e.g., class files) and additional \emph{package metadata} (e.g., \texttt{MANIFEST.MF} files) are archived in a \emph{deliverable package} for distribution. In \textsf{Java}-based systems, deliverable packages are distributed as a deployable \textsf{Java} application (e.g., a \texttt{war} file) or a \textsf{Java} library (e.g., a \texttt{jar} file). Similarly, the deliverable packages of \textsf{C/C++}-based systems are distributed as platform-dependent executable files (e.g., in \textsf{ELF} format) or shared libraries (e.g., \texttt{so} files).

\subsection{Existing approaches towards producing verifiable builds}
\label{sec:background-motivation:subsec:existing-approaches-verifiable-build}

Three main approaches can represent prior efforts towards verifiable builds. We refer to the first approach as ``controlling'', which comprises a mechanism that intercepts non-deterministic build instructions at runtime and replaces the returning value of these instructions with pre-defined deterministic values~\cite{leija:2020:repro_containers,ShiTSE2021}. The second approach is called ``remediation'' and comprises directly modifying non-deterministic instructions in the source code or build scripts. We refer to the third approach as ``interpretation'', which provides legitimate explanations about non-equivalence in generated build artifacts. Optionally, additional post-processing step(s) can be introduced to demonstrate the correctness of the explanations.

Prior research studies have been conducted to ensure build verifiability of \textsf{C/C++}-based systems. Carnavalet et al.~\cite{Carnavalet:2014} observe several related challenges to build verifiability of OSS systems. The authors manually identify and explain a set of sources of non-determinism in security-critical OSS systems. Ren et al.~\cite{Ren:2018,Ren:2019} adopt an automated build profiling technique to identify accountable instructions for introducing non-equivalences in the built packages. Our prior work~\cite{ShiTSE2021} proposes a unified process and a toolkit to produce verifiable builds of \textsf{C/C++} applications. Results show that the controlling mechanism implemented by our toolkit can mitigate most of the sources of non-determinism in both large-scale commercial systems and OSS systems. 
Leija et al.~\cite{Leija:2020} proposes a reproducible container, which can execute system calls in a deterministic way to eliminate sources of non-determinism from the build environment. Reproducible-Build~\cite{ReproBuild:2021} is a community-based effort to document the best practices and relevant tools for checking and verifying build reproducibility. It mainly focuses on \textsf{C/C++} systems (e.g., packages of the \textsf{Debian} distribution of \textsf{Linux}) and highlights that producing reproducible builds of \textsf{Java} systems is challenging.

\subsection{The current state of build verifiability for \textsf{Java}-based systems}
\label{sec:background:subsec:verifiable-builds-java-systems}

There are different tools for mitigating some sources of non-determinism and verifying deliverable packages for \textsf{Java}-based systems~\cite{jfrogRepro,gradleRepro,maveReproPlugin, zlikaMavenPlugin}. Each of these tools addresses one or more of the following three sources non-determinism: 
a) \textit{timestamps}: \texttt{jar} and configuration files (e.g., \texttt{pom.xml}) contain timestamps that are either replaced by pre-defined values or stripped off~\cite{repbuildmaventimestamp}, b) \textit{file order}: depending on the build process, packaged files in a \texttt{jar} file can have different order. Such packages files are then sorted after the build process is finished~\cite{reprojvm}, c) \textit{metadata on \texttt{manifest.mf} files}: user names and tooling version that are recorded in manifest files are stripped off~\cite{cipherkit}. The aforementioned solutions are natively supported by the major automated build tools (namely \textsf{Maven} and \textsf{Gradle}).

However, two main limitations render these solutions unsuitable for verifying the build of \textsf{Java}-based systems in an industrial setting: 
(1) \emph{Limited tool capability}: our experience on building industrial \textsf{Java}-based systems shows that there are several sources of non-determinism not covered by the provided solutions, such as sorting of symbol tables in the generated \texttt{jar} files and other non-equivalences introduced by specific tools (e.g., the \textsf{Jasper} compiler); and (2) \emph{Complex installation and configuration processes}: since none of the existing solutions supports all the usage scenarios and build tools, one has to install and configure all of them to provide a general solution used in the industrial context. This characteristic requires huge manual effort and cannot scale to various build environments and settings, typically needed by industrial systems. These limitations motivated us to develop a new approach to produce verifiable builds for industrial \textsf{Java}-based systems that, compared to the native solutions offered by automated build tools, is more flexible, extensible, and generalizable. We will cover the details of our approach in the next section.
\section{Our approach}
\label{sec:approach}



\begin{figure}
	\includegraphics[scale=0.4]{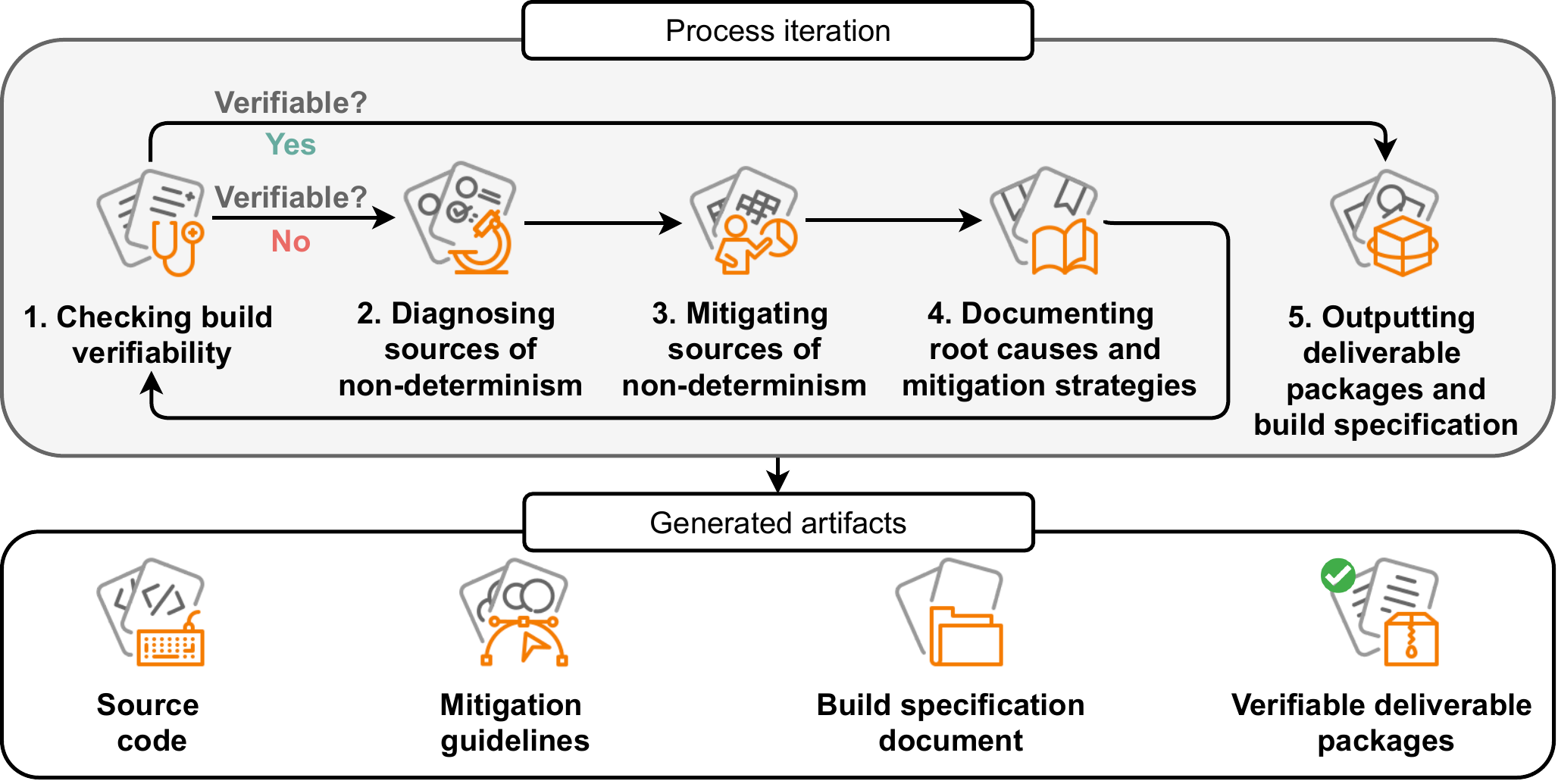}
	\centering
	\caption{An overview of our approach.}
	\label{fig:our-approach}
\end{figure}

As shown in Figure~\ref{fig:our-approach}, our approach consists of five phases.
(1) During the \textit{Checking build verifiability} phase, we prepare the build environment and invoke the build process. Then we check if the deliverable package is verifiable. 
(2) During the \textit{Diagnosing sources of non-determinism} phase, we study from existing literature and tool documentation to diagnose the sources of non-determinism in the deliverable package.
(3) During the \textit{Mitigating sources of non-determinism} phase, we configure our developed tools to control and interpret various sources of non-determinism.
(4) During the \textit{Documenting root causes and mitigation strategies} phase, we move back to Phase 1 to recheck the deliverable package. The process is repeated until the deliverable package is successfully verified. Then we document the root causes of non-determinism and update the corresponding mitigation strategies. Then in the (5) \textit{Outputting the deliverable package and build specifications} phase, we output the verified deliverable package along with the build specifications, which clearly describe the build environment and setup. 

To ease explanation, in the rest of the section, we will describe our approach using a running example, which is a \textsf{Java}-based system \textsf{Foo} consisting of two source code files, \texttt{Bar.java} and  \texttt{Baz.java}, and a \textsf{Maven} configuration file \texttt{pom.xml}. After build, it will generate a deliverable package~\texttt{Foo.jar}, which contains the following build artifacts: two \texttt{class} files (\texttt{Bar.class} and \texttt{Baz.class}) and three configuration files (\texttt{MANIFEST.MF}, \texttt{pom.properties}, and \texttt{pom.xml}). 

\noindent \textbf{Phase 1 - Checking build verifiability.}
The objective of this phase is to check the verifiability of the deliverable package generated from our build process.
We follow the same setup previously adopted in \textsf{C/C++}-based systems~\cite{ShiTSE2021}, where we use the same build environment, build specifications, and build commands to start two build processes to produce deliverable packages. This phase is further broken down into three steps:

\noindent \textit{Step 1 - Collecting build specifications:} In this step, we collect build information (e.g., \textsf{JDK} version, build tools, and dependencies) and record it in build specifications documents.

\noindent \textit{Step 2 - Setting up the build environment:} In this step, we prepare the build environment according to the build specifications. Typically, we encapsulate the build environment in a container (e.g., \textsf{docker}) or a virtual machine (VM) to ensure that the build environment is consistent. In our running example, we use the same build environment throughout (\textsf{docker} with \textsf{Ubuntu 18.04 LTS}), \textsf{JDK 1.8.0\_111}, and \textsf{Maven 3.6.0}).

\noindent \textit{Step 3 - Invoking the build process and checking build verifiability:} In this step, we carry out the build processes following the specified procedure. Two repeated build processes are invoked in the same environment with the same setup. 
To ease explanation, we refer to the resulting deliverable package of the first build process as $DP_{_{1}}$ and the resulting deliverable package of the second build process as $DP_{_{2}}$. 
We compare the SHA1-checksums of $DP_{_{1}}$ and $DP_{_{2}}$. If they are identical, we consider that the build is verifiable and move on to Phase 5.
If not, the build is not verifiable. Further inspection will be carried out in the next phase to diagnose sources of non-determinism. 

\noindent \textbf{Phase 2 - Diagnosing sources of non-determinism.} The objective of this phase is to identify the root cause of the non-equivalences in the two deliverable packages generated in Phase 1.
This phase is conducted manually and consists of the following two steps:

\noindent \textit{Step 1 - Studying existing research literature and tool documentation:} As there are no prior research studies focusing on \textsf{Java}-based systems, in this step, we first search on existing research work on build verifiability for \textsf{C/C++}-based systems~\cite{Ren:2018,Ren:2019,ShiTSE2021,Carnavalet:2014}. Then we collect various tools for producing verifiable builds for \textsf{Java}-based systems in the wild~\cite{reprojvm,mavenReproGuide,gradleRepro} and summarize the objectives and the solutions from these tools by studying their documentation. The collected knowledge will benefit us in the diagnosis process.

\noindent \textit{Step 2 - Comparing build artifacts:} In this step, we first unpack two non-equivalent deliverable packages $DP_{_{1}}$ and $DP_{_{2}}$ to extract two lists of build artifacts. The build artifacts usually consist of a set of \texttt{class} files and a set of text-based files. We apply diffing tools (e.g., \texttt{diffoscope}) to compare the build artifacts that have the same path and name. Since the \texttt{class} files are in the bytecode format, we first apply the \texttt{javap} command to decompile them into text-based representation. Then we compare the text-based representations line by line to examine the differences. For text-based files such as \texttt{MANIFEST.MF}, we directly examine the non-equivalences.
We also check the orders and file properties (e.g., created time) of build artifacts embedded in the deliverable packages as it also introduces non-equivalences in the deliverable package. We cross-check the identified non-equivalence and collected knowledge to summarize the root cause of these non-equivalences.

In our running example, after unpacking the deliverable package, five build artifacts are extracted.
Four build artifacts: \texttt{Bar.class}, \texttt{Baz.class}, \texttt{MANIFEST.MF}, and \texttt{pom.xml} are equivalent, while \texttt{pom.\\ properties} is not equivalent.
Listing~\ref{listing:example_diff} shows the non-equivalences that exist in the auto-generated messages, which contain timestamps. The timestamp differences are caused by the build environment, as two build processes are invoked at different times. This pattern of non-equivalence is further explained in Section~\ref{sec:evaluation:pattern} as \textsf{[P1]}.
In addition, the order of the build artifacts in the deliverable package is not deterministic. In $DP_{_{1}}$, the file \texttt{Bar.class} is listed before \texttt{Baz.class}, and it is the other way around in $DP_{_{2}}$. This is caused by the multi-threaded compilation of the \textsf{Java} compiler. This pattern of non-equivalence is further explained in Section~\ref{sec:evaluation:pattern} as \textsf{[P11]}.

\begin{lstlisting}[basicstyle=\ttfamily\small, frame=lines, label=listing:example_diff, caption=Example differences of timestamp in \texttt{pom.properties}.]
# Generated by Maven
- # Sun Sep 18 22:43:23 EDT 2021 
+ # Sun Sep 18 22:45:35 EDT 2021
\end{lstlisting}



\noindent \textbf{Phase 3 - Mitigating sources of non-determinism.} The objective of this phase is to leverage automated techniques to mitigate sources of non-determinism. 
This phase consists of two steps, which are automatically invoked during the compiling and linking, and the packaging phases of the build process, respectively.

\noindent \textit{Step 1 - Applying bytecode instrumentation to control sources of non-determinism:} In this step, sources of non-determinism are controlled by dynamically altering the behavior of build tools via bytecode instrumentation. 
Modern build tools (e.g., \textsf{Ant}, \textsf{Maven}, and \textsf{Gradle}) for \textsf{Java}-based systems are also implemented in \textsf{Java}. Hence, these build tools must first be loaded by \textsf{JVM} to start the build process. 
By default, the build tools write metadata such as timestamps into various build artifacts. Some metadata is non-deterministic and cannot be easily mitigated.
Hence, we develop a technique called \textsf{JavaBEPEnv}, which includes a custom \textsf{Java Agent} program. It leverages the \textsf{Java Instrument API} to modify the bytecode of build tools dynamically. \textsf{Java Instrument API} is a set of APIs supported by \textsf{JVM} to instrument bytecode when it is being loaded in the \textsf{JVM}.
To control the sources of non-determinism, \textsf{JavaBEPEnv} replaces the non-determinism introducing methods (e.g., \texttt{currentTimeMillis()}) with customized methods.
These customized methods have the same method signatures as the non-determinism introducing methods, but they only return deterministic information (e.g., a fixed timestamp) to keep the outputted information consistent.
\textsf{JavaBEPEnv} can be configured to be attached to \textsf{JVM} when the build process starts and control the non-deterministic behavior automatically.

\noindent \textit{Step 2 - Applying post-processing to interpret sources of non-determinism:} In the build process of \textsf{Java}-based systems, many non-equivalences are caused by different phases of the process (e.g., compiling and linking, and packaging), which cannot be easily controlled. Post-processing the build artifacts by rules could explain the non-equivalences and verify they are not malicious.
 
Interpreting sources of non-determinism means that an open and transparent technique will be automatically applied to build artifacts to eliminate non-equivalences. The technique should be transparent to third-party stakeholders who want to verify the build independently. We implement such a technique called \textsf{JavaBEPFix}, including rules like: (1) it leverages the \textsf{Byte Code Engineering Library} (\textsf{BCEL}) to modify the non-equivalent \texttt{class} files. \textsf{BCEL} is an open-source library to analyze, transform, and manipulate class files. With \textsf{BCEL}, \textsf{JavaBEPFix} is able to interpret many sources of non-determinism that cause non-equivalent \texttt{class} files, such as \textsf{[P6]} Constant pool and \textsf{[P7]} Temporary variables (details shown in Section~\ref{sec:evaluation:pattern}). (2) It automatically unpacks the deliverable package, sorts the build artifacts in a deterministic order based on predefined configuration (e.g., sort the files by name), and repacks the build artifacts into a post-processed deliverable package. (3) It also leverages the standard \textsf{java.nio API} to keep the creation time of build artifacts consistent. 
Other sources of non-determinism that can be interpreted by \textsf{JavaBEPFix} are discussed in Section~\ref{sec:evaluation:pattern}.
 
Similar to \textsf{C/C++}-based systems, to mitigate some sources of non-determinism in the deliverable packages of \textsf{Java}-based systems, we also need to leverage the remediation strategy (e.g., editing source code or configuration files, or upgrading dependencies). We will not mention the details here due to page limitations. Further details about how remediation is applied to produce verifiable builds can be found in our prior work~\cite{ShiTSE2021}.

In our running example, both \textsf{JavaBEPEnv} and \textsf{JavaBEPFix} are enabled during the build process. As a result, the timestamp generated in \texttt{pom.properties} files will always be the same, as the methods that generate timestamps are intercepted and transformed. All the build artifacts are automatically sorted and repacked into a deliverable package by \textsf{JavaBEPFix}.

\noindent \textbf{Phase 4 - Documenting root causes and mitigation strategies.} The objective of this phase is to document root causes of non-determinism and the corresponding mitigation strategies.
For each newly discovered source of non-determinism, we document its root cause and summarize it into a specific category. We also describe the recommended mitigation strategies. Such documentation is beneficial for producing verifiable builds in \textsf{Java}-based systems, as it can be reused whenever a documented root cause is identified. Please refer to Section~\ref{sec:evaluation:pattern} for a detailed documentation of various patterns of non-determinism in \textsf{Java}-based systems.
After this phase, we move back to Phase 1 and check if the new deliverable package after mitigation is verifiable.

In our running example, we document the patterns of timestamp and entries in deliverable packages. The root cause of non-equivalent timestamps is the build environment. The mitigation strategy is to control the timestamp using \texttt{JavaBEPEnv}.
The root cause of randomly ordered entries in deliverable packages is due to multi-threaded compiling. The mitigation strategy is to interpret the non-equivalences by sorting the built artifacts.



\noindent \textbf{Phase 5 - Outputting deliverable packages and build specifications.}
This phase begins once we deem the deliverable package as verifiable, after Phase 1.
The objective of this phase is to output the verifiable deliverable packages along with build specifications. 

The build specifications consist of three parts: (1) the build environment; (2) the build commands; and (3) the additional operations applied on the build artifacts for build verifiability.
The build environment could either be a \textsf{docker} file (if the build is started within a container), a VM image, or a detailed description of the host operating system (OS), and dependent libraries (e.g., the detailed versions of \textsf{JDK}), and so on. The build commands include the exact instructions to start the build process. The applied mitigation strategies are also documented. 
Independent builders can leverage the build specifications to verify the deliverable packages. Similar to our prior work~\cite{ShiTSE2021}, the outputs of this phase are provided to independent agencies for security auditing if needed.

We have included all the above information in our running example and delivered them to third-party auditing agencies. They acknowledged the build is verifiable.

\section{Case Study}
\label{sec:evaluation}

In this section, we present the evaluation of applying our approach on \textsf{Java}-based systems. In the past few years, we have applied our approach on various systems ranging from OSS to commercial systems within \textsf{Huawei}. These systems are from different application domains such as server-based systems, middleware libraries, and mobile apps. 
Due to confidentiality, we cannot directly discuss the evaluation details on our commercial systems. Hence, to demonstrate the effectiveness of our approach, we have applied our approach on various representative \textsf{Java}-based OSS systems. The case study setup and the evaluation results are described in Section~\ref{sec:evaluation:setup}.
Section~\ref{sec:evaluation:pattern} shares the sources of non-determinism and our proposed mitigation strategies based on our experience on applying our approach on thousands of \textsf{Java}-based commercial and open source systems.

\subsection{Performance Evaluation}
\label{sec:evaluation:setup}

In this section, we describe our case study setup (Section~\ref{sec:evaluation:subsec:performance-evaluation:subsubsec:case-study-setup}) and evaluation results (Section~\ref{sec:evaluation:quant}).

\subsubsection{Case Study Setup}
\label{sec:evaluation:subsec:performance-evaluation:subsubsec:case-study-setup}

Here we describe how to set up our case study on two different datasets of OSS projects. We first present the setup of OSS projects from \textsf{Reproducible Central}~\cite{mavenrebuild}. Next, we present the setup of OSS projects that are commonly adopted as dependencies within \textsf{Huawei.}

	
\noindent \textbf{OSSs from \textsf{Reproducible Central}.} \textsf{Reproducible Builds}~\cite{ReproBuild:2021} presents a collection of efforts on producing reproducible builds for \textsf{C/C++}-based systems. Recently, efforts towards reproducible builds for \textsf{Java}-based systems are also included~\cite{reprojvm}. \textsf{Reproducible Central}~\cite{mavenrebuild} is part of \textsf{Reproducible Build} efforts, which rebuilds open source \textsf{Java}-based systems and compare the deliverable packages with the stored ones in \textsf{Maven Central}. As of September 4, 2021, it contains 391 releases of 118 projects. Among them, 112 releases in 46 projects (the builds of 39\% projects) cannot be reproduced or verified. For demonstration purposes, we only focus on the build verifiability of the most recent releases of these 46 projects which are not verified. Table~\ref{tab:quanti} shows the basic information of these projects. The sizes of these 46 projects range from 108 lines to 578,998 lines, and they contain from 1 to 2,310 files. Examples of these 46 projects include \textsf{dubbo} and \textsf{kubernetes-client}. For brevity, we call these projects as \textsf{Reproducible Central} (RC) projects.
	
To examine if our approach can produce verifiable builds for RC projects, we follow the projects' build specification and build commands in a fresh \textsf{docker} environment as described in Section~\ref{sec:approach}. The main focus is to check if we can produce verifiable deliverable packages. Note that the deliverable packages stored in \textsf{Maven Central} were not built with our approach, hence many of the non-determinism have not been mitigated (e.g., not controlling the timestamp differences). To demonstrate the effectiveness of our approach, we have to compare the two deliverable packages built in our local environment, instead of comparing the deliverable packages against the ones in the central repository.

\noindent \textbf{Commonly adopted OSSs within \textsf{Huawei}.} Since the selected projects from \textsf{Reproducible Central} are generally of a smaller scale, to ensure generalizability, we have also selected the most recent releases of 13 open source projects which are widely adopted within \textsf{Huawei}. As shown in Table~\ref{tab:quanti}, the sizes of these commercially adopted projects range from 4,710 to 700,668 lines of code and have 51 to 7,540 source files. Examples of these 13 projects include \textsf{Spring Framework} and \textsf{SLF4J}. For brevity, we call these projects Commercially Adopted (CA) projects. In a similar setup as RC projects above, we also build the CA projects locally twice using the same build specification and build environments. Then, we verify the resulting deliverable packages.


\begin{table}[]
	\caption{Evaluation results after applying our approach. BVP represents build verifiable projects.}
	\footnotesize
	\centering
	\begin{tabular}{lrrrrrrr}
		\toprule
		\multirow{2}{*}{\textbf{Dataset}}               & \multirow{2}{*}{\textbf{Projects}} & \multicolumn{2}{c}{\textbf{BVP}} & \multicolumn{2}{c}{\textbf{SLOC}} & \multicolumn{2}{c}{\textbf{Files}} \\
		\cmidrule(lr){3-4} \cmidrule(lr){5-6} \cmidrule(lr){7-8}
		&& \textbf{Before} & \textbf{After} & \textbf{Min} & \textbf{Max} & \textbf{Min} & \textbf{Max} \\
		\midrule
		RC & 46    & 0 (0\%)        & 42  (91\%)           & 108 & 578,998  & 1 & 2,310  \\
		CA & 13    & 0 (0\%)        & 13  (100\%)          & 4,710 &  700,668 &  51 & 7,540 \\
		\bottomrule           
	\end{tabular}
	\label{tab:quanti}
\end{table}

\subsubsection{Case Study Results}
\label{sec:evaluation:quant}

In this section, we report the evaluation results of 46 RC projects and 13 CA projects. Table~\ref{tab:quanti} shows our evaluation results. For the build of 46 RC projects, before applying our approach, none of them are verifiable. After applying our approach, 42 (91\%) of them are successfully verified. The build of four RC projects failed to be fully verified due to additional non-deterministic APIs from third party libraries.

These four projects leverage third-party libraries to generate \textsf{Java} source code files, \textsf{XML} files, and index files. For example, \texttt{org.apache.royale.compiler} from \textsf{Apache Royale} uses \textsf{JFlex}, a lexical analyzer generator which can generate \textsf{Java} programs based on specifications. A set of \textsf{Java} source code files is generated for later use in the build process. The auto-generated source code files have differences in the comments (e.g., randomly sorted documentation for parameters used in a method), which lead to the non-equivalences in the deliverable packages. For such types of non-equivalences, we plan first to identify and locate the non-determinism introducing methods. Then we will extend the current implementation of \texttt{JavaBEPEnv} to dynamically instrument and alter the existing behavior of these methods. In particular, for the non-determinism introducing methods, we intercept them by defining custom methods with the same method signatures through \textsf{Java Instrumentation API}, and implement custom program logic to avoid non-determinism.

	
	

For the 13 CA projects, before applying our approach, none of them achieve build verifiability. After applying our approach, the build of all 13 CA projects can be successfully verified. The evaluation results on both RC and CA projects demonstrate the effectiveness of our approach towards build verifiability for \textsf{Java}-based systems.

\subsection{Our Mitigation Guidelines}
\label{sec:evaluation:pattern}


\lstset{
  language=Java,
  basicstyle=\scriptsize\ttfamily,
  breaklines=true,
  escapeinside={@:}{:@}
}

\noindent
\begin{table*}
\caption{Our mitigation guideline}
\footnotesize
\begin{tabular}{p{1.2cm}p{1.5cm}p{2.5cm}p{1.5cm}p{0.8cm}p{7.6cm}}
\toprule
\textbf{Root cause}  &  \textbf{Name}  & \textbf{Description} & \textbf{Strategy} & \textbf{Java-specific} & \textbf{Example} \\
\midrule
\multirow[t]{4}{1cm}{[RC1] Environment} &
[P1] Timestamp &
Time related information was written into files by build tools or embedded in the file properties. &
Control or Interpretation &
No &
\adjustbox{valign=t}{
\begin{lstlisting}
META-INF/vault/properties.xml @:\hfill \textsf{(wcm-caconfig-editor-1.8.0)}:@
- <entry key="created">2021-01-17T13:47:15.000Z</entry>
+ <entry key="created">2021-01-17T13:46:49.000Z</entry>
\end{lstlisting}} \\

\cmidrule{2-6}

&
[P2] JDK version &
JDK versions written into \texttt{MANIFEST.MF} &
Control &
Yes &
\adjustbox{valign=t}{
\begin{lstlisting}
META-INF/MANIFEST.MF @:\hfill \texttt{(io.dropwizard.metrics}@\textsf{metrics-servlets-4.1.22)}:@
- Build-Jdk: 1.8.0_292
+ Build-Jdk: 1.8.0_275
\end{lstlisting}} \\

\cmidrule{2-6}

&
[P3] Git information &
Git related information written into \texttt{git.json} and packaged in final artifact. &
Control &
No &
\adjustbox{valign=t}{
\begin{lstlisting}[boxpos=t]
classes/git.json @:\hfill \textsf{(ladapchai-0.8.1)}:@
- "git.local.branch.ahead": "0"
+ "git.local.branch.ahead": "NO_REMOTE"
\end{lstlisting}} \\

\cmidrule{2-6}

&
[P4] User information &
Users who invoked the build process written into \texttt{MANIFEST.MF}. &
Remediation or Control &
Yes &
\adjustbox{valign=t}{
\begin{lstlisting}[boxpos=t]
META-INF/MANIFEST.MF @:\hfill \texttt{(io.dropwizard.metrics}@\textsf{metrics-servlets-4.2.1)}:@
- Build-By: runner
+ Build-By: ?
\end{lstlisting}} \\

\midrule

\multirow[t]{6}{1cm}{[RC2] JDK} &
[P5] LineNumberTable &
Non-deterministic LineNumber Table generated by javac by default. &
Remediation &
Yes &
\adjustbox{valign=t}{
\begin{lstlisting}[boxpos=t]
io/fabric8/maven/docker/HelpMojo.class @:\hfill \textsf{(docker-maven-plugin-0.36.1)}:@
LineNumberTable:
- Line 29:0
+ Line 28:0
\end{lstlisting}} \\

\cmidrule{2-6}

&
[P6] Constant Pool &
Redundant/randomly ordered elements in Constant pool. &
Interpretation &
Yes &
\adjustbox{valign=t}{
\begin{lstlisting}[boxpos=t]
io/fabric8/maven/docker/HelpMojo.class @:\hfill \textsf{(docker-maven-plugin-0.36.1)}:@
- #12 = Methodref #160.#279 // java/io/InputStream.close:()V
+ #91 = Methodref #88.#90 // java/io/InputStream.close:()V
\end{lstlisting}} \\

\cmidrule{2-6}

&
[P7] Temporary variables &
The temporary variables have different assigned IDs. &
Interpretation &
Yes &
\adjustbox{valign=t}{
\begin{lstlisting}[boxpos=t]
ClassA.class    ClassA.class @:\hfill \textsf{(Internal project)}:@
- astore 15     + astore 13
- aload 14      + aload 12
\end{lstlisting}} \\

\cmidrule{2-6}

&
[P8] Javadoc &
Javadoc entries randomly sorted due to JDK bug. &
Control &
Yes &
\adjustbox{valign=t}{
\begin{lstlisting}[boxpos=t]
  @:\hfill (\textsf{JDK-8013887}/\textsf{Internal project}):@
- com.sun.source.tree       + com.sun.source.doctree
- com.sun.source.doctree    + com.sun.source.util
- com.sun.source.util       + com.sun.source.tree
\end{lstlisting}} \\

\cmidrule{2-6}

&
[P9] Inner class order &
The order of inner classes is non-deterministic. &
Interpretation &
Yes &
\adjustbox{valign=t}{
\begin{lstlisting}[boxpos=t]
InnerClasses: @:\hfill \textsf{(Internal project)}:@
- public static #160= #518 of #517; //Foo=class A/B/C/...
- public static #189= #520 of #519; //Bar=class A/B/C/...
public static #162= #650 of #657; //Baz=class A/B/C/...
+ public static #160= #518 of #517; //Foo=class A/B/C/...
+ public static #189= #520 of #519; //Bar=class A/B/C/...
\end{lstlisting}} \\

\cmidrule{2-6}

&
[P10] Method order &
Methods in class files are randomly ordered. &
Interpretation &
Yes &
\adjustbox{valign=t}{
\begin{lstlisting}[boxpos=t]
@:\hfill \textsf{(Kubernetes-client-project-5.4.1)}:@
Io/fabric8/kubernetes/api/model/WatchEventFluent.class 
A withAuthInfoObject(final AuthInfo p0);
- A withAPIServiceObject(final APIService p0);
A withResourceRequirementObject(...);
+ A withAPIServiceObject(final APIService p0);
\end{lstlisting}} \\ 

\midrule

[RC3] Multi-thread &
[P11] Entries in deliverable packages &
Files packaged in archive files randomly sorted due to multithreading. &
Interpretation &
No &
\adjustbox{valign=t}{
\begin{lstlisting}[boxpos=t]
- ... META-INF/MANIFEST.MF @:\hfill \textsf{(liquibase-percona-4.3.1.jar)}:@
- ... META-INF/services/
+ ... META-INF/MANIFEST.MF
+ ... liquibase/
\end{lstlisting}} \\

\midrule


[RC4] Other tools &
[P12] Properties in files &
Properties in \texttt{MANIFEST.MF} are randomly ordered. &
Interpretation &
Yes &
\adjustbox{valign=t}{
\begin{lstlisting}[boxpos=t]
@:\hfill \textsf{(io.dropwizard.metrics:metrics-4.2.1)}:@
- Export-Package:com.codahale.metrics.health;uses:="com.codahale.metrics";version="4.2.1" (...) com.codahale.metrics.health.annotation;version="4.2.1"
+ Export-Package:com.codahale.metrics.health.annotation;version="4.2.1",com.codahale.metrics.health;uses:="com.codahale.metrics";version="4.2.1" (...)
\end{lstlisting}} \\

\cmidrule{2-6}

&
[P13] JSP compilation &
Different source code generated by \textsf{Jasper} due to cache option. &
Control &
Yes &
\adjustbox{valign=t}{
\begin{lstlisting}[boxpos=t]
JasperGeneratedFile.java @:\hfill \textsf{(Internal project)}:@ 
+ static { _jspx_dependants = new
+ java.util.HashMap<java.lang.String,java.lang.Long>(2);
+ _jspx_dependants.put("dep1.jar",Long.valueOf(1685L));
...}
\end{lstlisting}} \\

\midrule

[RC5] Compound effect &
[P14] Lambda expression &
Auto-generated methods for lambda expression during compilation are not consistent. &
Control &
Yes &
\adjustbox{valign=t}{
\begin{lstlisting}[boxpos=t]
- #25 = Methodref #4.#30 // L1.lambda$new$0:()V @:\hfill \textsf{(Internal project)}:@ 
+ #25 = Methodref #4.#30 // L1.lambda$new$1:()V
- #25 = Methodref #4.#30 // L2.lambda$new$1:()V
+ #25 = Methodref #4.#30 // L2.lambda$new$0:()V
\end{lstlisting}} \\
\bottomrule
\label{tab:mitigation-guidelines}
\end{tabular} 
\end{table*}

This section describes various patterns of sources of non-determinism and the corresponding mitigation strategies. Table~\ref{tab:mitigation-guidelines} presents the list of patterns. For each pattern, we include the root cause, the description, the mitigation strategy, the specificity, and one code example. There are a total of 14 patterns from five categories of root causes. 
Three types of strategies are applied to mitigate the sources of non-determinism: Control, Interpretation, and Remediation. Control includes using \textsf{JavaBEPEnv} to dynamically alter non-deterministic behaviors or ensuring the build environment is consistent. Interpretation involves using \textsf{JavaBEPFix} to post-process non-equivalent build artifacts. Remediation includes modifying source code, configurations, or upgrading \textsf{JDK} versions. In the remainder of this section, we describe each pattern in detail.


\subsubsection{[RC1] Environment}
The environmental factors refer to the build environment that should be documented in build specifications, including the types of host OS (e.g., \textsf{Windows} or \textsf{Linux}), and the dependent \textsf{JDK} versions. Without such documentation, many non-equivalences might be introduced to the build artifacts. Such non-determinism could usually be avoided if a \textsf{docker} or VM is provided for the build, except for timestamp-related non-determinism. Below we list the common patterns we find during our evaluation.

\noindent\textbf{[P1] Timestamp.} Build tools call \textsf{JVM}-level functions to retrieve the current timestamp. The timestamp is then written to build artifacts, causing non-equivalences across different build instances. In addition, generated files might contain time-related information, such as the creation time.  Table~\ref{tab:mitigation-guidelines} shows an example. During the build process of \texttt{wcm-caconfig-editor-1.8.0}, the timestamps are recorded in the \texttt{properties.xml} file, which causes the build artifacts to be non-equivalent.

\noindent \emph{Solution:}
Use \textsf{JavaBEPEnv} to replace the timestamp introducing function calls at \textsf{JVM}-level with customized functions. The customized functions return the pre-defined timestamp instead of real timestamp values, preventing non-deterministic information from being written into build artifacts. For the timestamps recorded in the file attributes, use \textsf{JavaBEPFix} to process all the files and assigning pre-defined timestamps to them.

\noindent\textbf{[P2]~\textsf{JDK} version.} 
The \textsf{JDK} version used in the build process is written into \texttt{MANIFEST.MF}. As shown in Table~\ref{tab:mitigation-guidelines}, during the build process of \textsf{dropwizard}~\cite{pattern:jdkversion}, the two build instances invokes two different \textsf{JDK} versions: 1.8.0\_292 and 1.8.0\_275, as the build specification does not record the exact \textsf{JDK} version while it simply notes \texttt{JDK1.8}. When independent builders try to verify the build, a different \textsf{JDK} version is recorded in the \texttt{MANIFEST.MF}, causing non-equivalences in the build artifact.

\noindent\emph{Solution:} Make sure the same \textsf{JDK} version is used during the build processes. The adopted \textsf{JDK} version should be documented in the build specification for future references as well.

\noindent\textbf{[P3]~\textsf{Git} information.}
Many software projects use \textsf{Git} as the VCS to manage the evolution and the maintenance of the projects. Some build processes record \textsf{Git} related information (e.g., the \texttt{commit ID} or the user who started the process) in the build artifacts. Such information might be non-deterministic if the build processes are started in different environments. As shown in Table~\ref{tab:mitigation-guidelines}, one of the build environment for \texttt{ladapchai-0.8.1.jar} does not configure the remote repository~\cite{pattern:git}, causing the field \texttt{git.local.branch.ahead} to be non-equivalent. 

\noindent\emph{Solution:} Make sure the same \textsf{Git} setup is used during the build processes. The \textsf{Git} setup information should be documented in the build specification for future references as well. Recommend to use the same build environment (e.g., VMs or containers) to ensure the consistency of the build environment.

\noindent\textbf{[P4] User information.}
The \texttt{user ID} of the user who invokes the build process can be written into \texttt{MANIFEST.MF} file. As shown in Table~\ref{tab:mitigation-guidelines}, the user information is recorded in the \texttt{Build-By} field in \texttt{MANIFEST.MF} during the build process for \texttt{metrics-servlets-4.2.1}. As two build processes can be conducted in different environments, the user information can also be different.

\noindent\emph{Solution:} Modifying the build configuration can mitigate this issue. A configuration field \texttt{Built-by} under \texttt{manifestEntries} can be set with a consistent name to avoid non-deterministic \texttt{user ID}s. Alternatively, build within consistent environments (similar to \textsf{[P2]} and \textsf{[P3]}) can also mitigate this source of non-determinism.

\subsubsection{[RC2] JDK}
Some non-deterministic behavior is caused by \textsf{JDK} during the build process. Below we list the common patterns that we find during our evaluation.

\noindent\textbf{[P5]~\textsf{LineNumberTable}.}
\texttt{LineNumberTable} is an optional attribute that represents the relation between source code and bytecode. It can vary during the compiling phase. Table~\ref{tab:mitigation-guidelines} shows an example. During the build process for \texttt{docker-maven-plugin}~\cite{pattern:linenumbertable}, the values \texttt{LineNumberTable} are different in the \texttt{HelpMojo.class} file.

\noindent\emph{Solution:} Modify the build configuration to mitigate this issue.
For example, a configuration parameter \texttt{-g:none} can be added with \texttt{javac} to prevent \texttt{LineNumberTable} from being written into the bytecode. If the build process is started by \textsf{Maven}, we can also disable the generation of \texttt{LineNumberTable} by adding such configuration in the \texttt{pom.xml}.

\noindent\textbf{[P6] Constant pool.}
The constant pool is a data structure inside \texttt{class} files. It records the symbolic references that \textsf{JVM} uses to link with the actual contents of variables, methods, interfaces, etc. We find that the constant pool might contain duplicated elements. The indices of the duplicated elements are used in a non-deterministic way when these elements are referenced.
Furthermore, the constant pool might be randomly ordered accross two build instances, causing the indices to be different. Table~\ref{tab:mitigation-guidelines} shows an example of this pattern. During two build instances of \texttt{docker-maven-plugin}~\cite{pattern:constantpool}, the indices of the reference to the \texttt{close} method are recorded as 12 and 91, respectively. 

\noindent\emph{Solution:} Use \textsf{JavaBEPFix} to mitigate this issue. In particular, it will post-process the \texttt{class} files by deduplicating the constant pool and then sort it in a deterministic order.

\noindent\textbf{[P7] Temporary variables.}
Temporary variables are variables with a short lifetime. For example, a return statement \texttt{return (a+b);} would create a temporary variable when compiled to bytecode. Such variables will be assigned with a temporary \texttt{ID} by the compiler, which is used for instructions such as \texttt{astore} and \texttt{aload}. We find the same build process could yield different \texttt{ID}s for the temporary variables. The example shown in Table~\ref{tab:mitigation-guidelines} is adapted from our internal project. At the same locations of a class file, the \texttt{ID}s of the temporary variables are different in two build instances.  

\noindent\emph{Solution:} Use \textsf{JavaBEPFix} to mitigate this issue. In particular, it will automatically post-process the \texttt{class} files to reassign temporary variables with deterministic \texttt{ID}s.

\noindent\textbf{[P8]~\textsf{JavaDoc}.}
Lower versions of \textsf{JDK} can cause entries in the \textsf{JavaDoc} being randomly sorted~\cite{pattern:javadoc}. 
The example shown in Table~\ref{tab:mitigation-guidelines} is adapted from the issue report \texttt{JDK-8013887}~\cite{pattern:javadoc}, where the three \textsf{JavaDoc} entries have different orders during two identical build processes. We found such non-determinism exists in \textsf{Huawei}'s internal projects.

\noindent\emph{Solution:} Upgrade the \textsf{JDK} version to be higher than or equal to \texttt{1.8\_b105} to mitigate this issue.  

\noindent\textbf{[P9] Inner class order.}
Inner classes are classes that are defined inside another class. When the source code of the class that contains inner classes is compiled to bytecode, the list of inner classes is listed in the bytecode. As shown in Table~\ref{tab:mitigation-guidelines}, three inner classes, \texttt{Foo}, \texttt{Bar}, and \texttt{Baz} are listed. However, the order of the inner classes in the list is non-deterministic.

\noindent\emph{Solution:} Use \textsf{JavaBEPFix} to mitigate this issue. In particular, it will automatically post-process the \texttt{class} files by sorting the inner classes in a deterministic way.

\noindent\textbf{[P10] Method order.}
The order of the compiled methods in the \texttt{class} files might be non-deterministic.
Table~\ref{tab:mitigation-guidelines} shows an example.
Across two build instances  of \texttt{kubernetes-client-project-5.4.1}, file \texttt{WatchEventFluent.class} is not equivalent. 
We find that the only difference is the order of the methods inside the class (e.g., the method \texttt{withAPIServiceObject} appears before or after the method \texttt{withResourceRequiremetnObject}).  

\noindent\emph{Solution:} Use \textsf{JavaBEPFix} to mitigate this issue. In particular, it will automatically post-process the \texttt{class} files by sorting the methods in a deterministic way.

\subsubsection{[RC3] Multi-thread} 
Multi-threaded compilation is widely adopted by modern build tools, as it can accelerate the build process. However, build artifacts might be generated in a non-deterministic order, causing non-equivalences in deliverable packages.

\noindent \textbf{[P11] Entries in deliverable packages.}
Each deliverable package contains a list of build artifacts, which are compiled and packaged in a multi-threaded manner. 
The sequence of these build artifacts in the deliverable package is not deterministic, as it depends on which thread execution finishes first.
As a consequence, although the contents of each build artifact are equivalent during two build instances, but the deliverable package as a whole is not.
As shown in Table~\ref{tab:mitigation-guidelines}, during the build processes of \texttt{liquibase-percona-4.3.1.jar}, the entries are not ordered the same, causing the resulting deliverable package to be different.

\noindent\emph{Solution:} Use \textsf{JavaBEPFix} to mitigate this issue. In particular, after the original deliverable package is generated, it first unpacks the deliverable package and then re-packages the build artifacts in a deterministic way (e.g., by name).

\subsubsection{[RC4] Other tools} 
Build processes of some \textsf{Java}-based systems depend on third-party tools or plugins. These tools and plugins might introduce non-determinism into build artifacts, causing non-equivalences in the deliverable packages.

\noindent\textbf{[P12] Properties in files.}
Some properties in \texttt{MANIFEST.MF} files might be randomly ordered.
For example, the \texttt{Export-Package} property records the packages that are visible outside the deliverable package. In the build of \texttt{dropwizrd-metrics-4.2.1.jar}, the list of \texttt{Export-Package} does not have the consistent sequence. Similar issue is identified with the \texttt{Private-Package} attribute. This is caused by a third-party build plugin tool.

\noindent\emph{Solution:} Use \textsf{JavaBEPFix} to mitigate this issue. In particular, it will automatically locate the non-deterministic file properties (e.g., \texttt{Private-Package} and \texttt{Export-Package}) in \texttt{MANIFEST.MF} file and sort the relevant properties. 

\noindent\textbf{[P13] JSP compilation.}
\texttt{JSP} is a Java-based technique to create dynamically generated webpages. The \texttt{JSP} files can be parsed to \textsf{Java} source code files so that they share all the APIs and functionalities provided by \texttt{JVM}. To do that, techniques such as \textsf{Tomcat Jasper} engine are applied. A source of non-determinism is identified during this process, as the caching option in the \textsf{Jasper} engine cause the generated source code to have non-equivalent static variable definitions as shown in Table~\ref{tab:mitigation-guidelines}.

\noindent\emph{Solution:} Use \textsf{JavaBEPEnv} can mitigate this issue. It sets the caching option to false and prevents the inconsistent information being generated through dynamic instrumentation.

\subsubsection{[RC5] Compound effects} 
Some non-equivalences are due to a combination of root causes such as multi-thread and JDK behavior.

\noindent\textbf{[P14] Lambda expressions.}
This pattern occurs when there are multiple lambda expressions in the source code. When compiling with lower versions of \textsf{JDK} in a multi-threaded setting, the \texttt{ID} of the same lambda expressions may be assigned in a different way~\cite{pattern:lambda}. As shown in Table~\ref{tab:mitigation-guidelines}, the \texttt{ID}s of the lambda expressions of the two files, \texttt{L1.java} and \texttt{L2.java}, are different when the two files are compiled in a different order.

\noindent\emph{Solution:} Upgrade \textsf{JDK} to a version that is newer than \texttt{jdk8-b44}.

%

\section{Discussion}
\label{sec:discussion}

In this section, we discuss how our approach can help on tackling related industrial challenges to build verifiability of \textsf{Java}-based systems and present some future research directions.

\noindent \textbf{Towards trustworthy software supply chains.} Software supply chain attacks explore the dependency relationships between different software components and target software systems. One form of supply chain attack is the injection of malicious code during the build process~\cite{marc:backstabber_knife:2020}, particularly by hijacking third-party libraries distributed through central repositories and linked to the built package. Verifiable build plays an important role in preventing this type of software supply chain attack. In particular, systems with a verifiable build can have their integrity jointly verified by independent builders that share checksums of the generated build artifacts, such that others can compare against the build artifacts that they produce themselves. For example, Lamb and Zacchiroli~\cite{LambIeeeSftw:2021} discuss how reproducible builds can help OSS users in establishing trust in distributed packages through package managers. 
Compared to deliverable packages built locally, those in the remote central repositories usually suffer from more types of non-equivalences due to inconsistent build environment, lack of automated techniques, and other related factors. Our approach proposes an important step to support a trustworthy software supply chain, as it helps developers to deploy verifiable build artifacts in central repositories.

\noindent \textbf{Comparison across different OS platforms.} Deliverable packages should be verifiable even when built across different OS platforms, as one major advantage of \texttt{Java} is the compatibility across different OSs.
We have conducted additional experiments on 13 CA projects on both \textsf{Linux} and \textsf{Windows} platforms.
Our results show that the deliverable packages are verifiable when built in \textsf{Linux} and \textsf{Windows} separately by applying our approach.
However, when comparing the deliverable packages between \textsf{Linux} and \textsf{Windows}, none of the build packages are equivalent.
This is due to the environmental differences between the two OS platforms.
Take the build processes for \texttt{Logback} as an example. The two OS platforms can have different users, which triggers the \texttt{[P4]} pattern. In addition, even using the same \textsf{JDK} versions, there are differences in the class files while building in \textsf{Linux} and \textsf{Windows}. 
Although one of the main advantages of \textsf{Java} is that software systems can be ``built once and run anywhere'', the verifiability of the build across different platforms is still not satisfied. This issue can be currently resolved by specifying the OS platforms as part of the build specifications or using pre-setup VMs. Such requirement is in accordance with the definition of build verifiability~\cite{ReproBuild:2021}.
However, mitigating the sources of non-determinism from different platforms remains to be an interesting piece of future work.

\noindent \textbf{Comparison among systems implemented in different programming languages.} Due to the different setups of interpreted vs. compiled programming languages, the mitigation strategies also differ in the following two aspects:
(a)\textit{Different approaches for same mitigation strategies:} Various prior work has been done towards build verifiability for \textsf{C/C++}-based systems~\cite{Ren:2018,Ren:2019,ShiTSE2021}. The strategy of control in our prior work~\cite{ShiTSE2021} is similar to \textsf{JavaBEPEnv} proposed in this paper. This mechanism intercepts non-determinism introducing functions (e.g., functions that return timestamp) and returns pre-defined values. However, the approach in~\cite{ShiTSE2021} cannot be directly used in \textsf{Java}-based systems. The control strategy of our prior work focuses on process level, where the system level functions are intercepted. However, in the build process of \textsf{Java}-based software systems, \textsf{JVM} is the only process that is instantiated. Furthermore, simply intercepting the \textsf{JVM} process will likely cause the congestion in the build process, as many multi-threaded operations in \textsf{JVM} is time-sensitive. (b) \textit{\textsf{Java}-specific patterns}: As shown in Table~\ref{tab:mitigation-guidelines}, there are three common patterns associated with non-equivalent build artifacts between \textsf{Java}- and \textsf{C/C++}-based systems. However, there are also eleven patterns that are unique for \textsf{Java}-based systems. For example, six patterns that are caused by \textsf{JDK} behavior are specific to \textsf{Java}-based systems. Other software systems that are built on top of \textsf{JVM} (e.g., \textsf{Kotlin} and \textsf{Scala}) may benefit from our approach and future work should investigate the build verifiability for systems written in \textsf{JVM}-supported programming languages.



\section{Threats to Validity}
\label{sec:threats-to-validity}

In this section, we present the threats to validity. 


\noindent\textbf{External Validity.} We try to be as comprehensive as we can in our case study by selecting 46 projects from \textsf{Reproducible Central}, and 13 projects from commercially adopted OSS systems. However, our approach and our mitigation strategies towards build verifiability may not cover all the build verifiability scenarios for \textsf{Java}-based systems. 
In addition, our approach and findings are only limited to \textsf{Java}-based systems and may not be applicable to software systems written in other programming languages.


\noindent\textbf{Internal Validity.} To avoid confounding factors, we ensure the build environment is consistent before we perform our experiments. Although \textsf{Java} is a programming language that is OS independent, we still only perform our experiments on the same OS platforms (either \textsf{Linux} or \textsf{Windows}) depending on the usage scenarios. This setup is in accordance with the definition of build verifiability~\cite{ReproBuild:2021}.


\noindent\textbf{Construct Validity.} To check if the build is verifiable, we track all the deliverable packages generated by the build process. For example, a build process generates 100 deliverable packages (e.g., 100 \texttt{jar} files). If there is at least one \texttt{jar} file which is not verifiable, we do not consider the build verifiable. Our approach is similar to the prior work in this area~\cite{Ren:2018,Ren:2019,ShiTSE2021}.



\section{Conclusions}
\label{sec:conclusion}

Build verifiability is essential for software security and trustworthiness. While various prior work has been done to ensure verifiable build for \textsf{C/C++} systems, there is a lack of systematic solution for \textsf{Java}-based systems. 
In this paper, we propose a systematic approach towards build verifiability in \textsf{Java}-based systems. 
Our approach includes a unified process and two main techniques: a tool \textsf{JavaBEPEnv}, which controls non-deterministic behavior from the build tools, and another tool \textsf{JavaBEPFix}, which interprets non-determinism by post-processing non-equivalences in build artifacts. Case studies show that among 59 OSSs which are not build verifiable, 55 (93\%) projects are now build verifiable by applying our approach. 
We also present a mitigation guideline, which includes all the sources of non-determinism we encountered and the corresponding mitigation strategies. Last, we discuss some challenges and provide some open research problems.

\bibliographystyle{ACM-Reference-Format}
\bibliography{seip2022}

\end{document}